\documentclass[aps,prb,a4paper,10pt,twocolumn,showpacs,floatfix,superscriptaddress,preprintnumbers,longbibliography]{revtex4-2}
\usepackage{float}
\usepackage{dcolumn,graphicx,color,booktabs,microtype,afterpage}
\usepackage{amssymb}
\usepackage{amsmath}
\usepackage[charter,greekuppercase=italicized]{mathdesign}
\usepackage{sidecap}
\usepackage[mathlines]{lineno}

\graphicspath{{./}{figure/}}
\renewcommand{\tablename}{Table}
\makeatletter\renewcommand{\fnum@figure}[1]{\figurename~\thefigure.~}\makeatother
\makeatletter\renewcommand{\fnum@table}[1]{\tablename~\thetable.}\makeatother

\newcount\hh \newcount\mm
\hh=\time \divide\hh by 60
\mm=\hh \multiply\mm by 60 \mm=-\mm
\advance\mm by \time
\def\now{\number\hh:\ifnum\mm<10{}0\fi\number\mm}

\usepackage[colorlinks,plainpages=false,linkcolor=blue,urlcolor=blue,citecolor=blue,pdfpagemode=UseNone,pdfstartview=FitBH]{hyperref}

\newcommand{\tcr}[1]{\textcolor{black}{#1}}

\begin{document}

\makeatletter\renewcommand{\ps@plain}{%
\def\@evenhead{\hfill\itshape\rightmark}%
\def\@oddhead{\itshape\leftmark\hfill}%
\renewcommand{\@evenfoot}{\hfill\small{--~\thepage~--}\hfill}%
\renewcommand{\@oddfoot}{\hfill\small{--~\thepage~--}\hfill}%
}\makeatother\pagestyle{plain}

\preprint{\textit{Preprint: \today, \now}} 
\title{\tcr{Nodeless} superconductivity in the centro- and noncentrosymmetric rhenium-boron superconductors}
%
\author{T.\ Shang}\email[Corresponding authors:\\]{tshang@phy.ecnu.edu.cn}
\affiliation{Key Laboratory of Polar Materials and Devices (MOE), School of Physics and Electronic Science, East China Normal University, Shanghai 200241, China}
%
%
\author{W.\ Xie}\thanks{Present address: Deutsches Elektronen-Synchrotron (DESY), Notkestra\ss{}e 85, 22607 Hamburg, Germany}
\affiliation{Center for Correlated Matter and Department of Physics, Zhejiang University, Hangzhou 310058, China}
\author{J.\ Z.\ Zhao}
\affiliation{Co-Innovation Center for New Energetic Materials, Southwest University of Science and Technology, Mianyang, 621010, China} 
\affiliation{Research Laboratory for Quantum Materials, Singapore University of Technology and Design, Singapore 487372, Singapore} 
\author{ Y.\ Chen}
\affiliation{Center for Correlated Matter and Department of Physics, Zhejiang University, Hangzhou 310058, China}
\author{D.\ J.\ Gawryluk}
\affiliation{Laboratory for Multiscale Materials Experiments, Paul Scherrer Institut, Villigen CH-5232, Switzerland}
\author{M.~Medarde}
\affiliation{Laboratory for Multiscale Materials Experiments, Paul Scherrer Institut, Villigen CH-5232, Switzerland}
\author{M.\ Shi}
\affiliation{Swiss Light Source, Paul Scherrer Institut, Villigen CH-5232, Switzerland}
\author{ H.\ Q.\ Yuan}
\affiliation{Center for Correlated Matter and Department of Physics, Zhejiang University, Hangzhou 310058, China}
\author{E.\ Pomjakushina}
\affiliation{Laboratory for Multiscale Materials Experiments, Paul Scherrer Institut, Villigen CH-5232, Switzerland}
\author{T.\ Shiroka}
\affiliation{Laboratory for Muon-Spin Spectroscopy, Paul Scherrer Institut, Villigen PSI, Switzerland}
\affiliation{Laboratorium f\"ur Festk\"orperphysik, ETH Z\"urich, CH-8093 Z\"urich, Switzerland}
\begin{abstract}
We report a comprehensive study of the centrosymmetric Re$_3$B and 
noncentrosymmetric Re$_7$B$_3$ superconductors. At a macroscopic level, 
their bulk superconductivity (SC), with $T_c$ = 5.1\,K (Re$_3$B) and 3.3\,K 
(Re$_7$B$_3$), was characterized via electrical-resistivity, magnetization, 
and heat-capacity measurements, while their microscopic superconducting 
properties were investigated by means of muon-spin rotation/relaxation ($\mu$SR).  
In both Re$_3$B and Re$_7$B$_3$ the low-$T$ zero-field electronic 
specific heat and the superfluid density (determined via tranverse-field $\mu$SR) 
suggest a nodeless SC. Both compounds exhibit 
\tcr{some} features 
of multigap SC,
as evidenced by temperature-dependent upper critical fields $H_\mathrm{c2}(T)$, 
as well as by electronic band-structure calculations.
The absence of spontaneous 
magnetic fields below the onset of SC, as determined from zero-field 
$\mu$SR measurements, indicates a preserved time-reversal symmetry in 
the superconducting state of both Re$_3$B and Re$_7$B$_3$.
Our results suggest that a lack of inversion 
symmetry and the accompanying antisymmetric spin-orbit coupling effects 
are not essential 
for the occurrence of multigap SC in these rhenium-boron compounds.
\end{abstract}

\maketitle\enlargethispage{3pt}

\section{\label{sec:Introduction}Introduction}\enlargethispage{8pt}
%
The possibility to host unconventional- and 
topological superconductivity (SC), or to act as systems where to 
realize the Majorana fermions~\cite{Bauer2012,Smidman2017,Ghosh2020b,Kim2018,Sun2015,Ali2014,Sato2009,Tanaka2010,Sato2017,Qi2011,Kallin2016}, has made noncentrosymmetric superconductors (NCSCs) one the 
most investigated families in recent times. In NCSCs, a lack of inversion 
symmetry implies that admixtures of spin-singlet and spin-triplet 
superconducting pairings are allowed~\cite{Bauer2012,Smidman2017,Ghosh2020b}. 
This sets the scene for a variety of exotic properties, such as, 
upper critical fields beyond the Pauli limit~\cite{Carnicom2018,Bauer2004}, 
nodes in the superconducting gap~\cite{yuan2006,nishiyama2007,bonalde2005CePt3Si,Shang2020}, 
or multigap SC~\cite{kuroiwa2008}. More interestingly, by using the 
muon-spin relaxation ($\mu$SR) technique, time-reversal symmetry (TRS) 
breaking
has been observed to occur in the superconducting state of selected 
weakly-correlated NCSCs. 
These include CaPtAs~\cite{Shang2020}, LaNiC$_2$~\cite{Hillier2009}, La$_7T_3$ ($T$ = transition metal)~\cite{Barker2015,Singh2018La7Rh3,Mayoh2021}, Zr$_3$Ir~\cite{Shang2020b}, and Re$T$~\cite{Singh2014,Singh2017,Shang2018a,Shang2018b}. 
Except for CaPtAs, where TRS breaking and superconducting gap nodes 
coexist below $T_c$~\cite{Shang2020,Xie2019}, in most other cases the superconducting 
properties resemble those of conventional superconductors, characterized 
by a fully opened energy gap. 
In general, the causes behind TRS breaking in these superconductors are 
not yet fully  
understood and remain an intriguing open question.

To clarify 
the issue, the $\alpha$-Mn-type Re$T$ superconductors 
have been widely studied and demonstrated to show a superconducting state 
with broken TRS~\cite{Singh2014,Singh2017,Shang2018a,Shang2018b}.
Our previous comparative $\mu$SR studies on Re-Mo alloys, covering four different crystal structures (including the noncentrosymmetric $\alpha$-Mn-type), reveal that the spontaneous magnetic fields occurring below $T_c$ were only observed in elementary rhenium and in Re$_{0.88}$Mo$_{0.12}$~\cite{Shang2018b,Shang2019,Shang2020ReMo}. By contrast, TRS was preserved in the Re-Mo alloys with a lower Re-content (below $\sim 88$\%), independent of their centro- or noncentrosymmetric crystal structures~\cite{Shang2020ReMo}. Since both elementary rhenium and Re$_{0.88}$Mo$_{0.12}$ adopt a simple centrosymmetric structure (hcp-Mg-type), this strongly
suggests that a noncentrosymmetric structure is not essential in realizing the TRS breaking in Re$T$ superconductors.  
The $\mu$SR results regarding the Re-Mo family, as well as 
other $\alpha$-Mn-type superconductors, e.g., Mg$_{10}$Ir$_{19}$B$_{16}$, Nb$_{0.5}$Os$_{0.5}$, Re$_3$W, and Re$_3$Ta~\cite{Acze2010,SinghNbOs,Biswas2012,Barker2018}, where TRS is preserved,  
clearly indicate that not only the Re presence, but also its amount is crucial for the appearance and the extent of TRS breaking in the Re$T$ superconductors. 
How these results can be understood within a more general framework 
clearly requires further investigations.

Rhenium-boron compounds represent another suitable 
candidate system for studying the TRS breaking effects in the family 
of Re-based superconductors. Indeed, upon slight changes of the Re/B ratio, 
both centrosymmetric Re$_3$B (C-Re$_3$B) and noncentrosymmetric Re$_7$B$_3$ 
(NC-Re$_7$B$_3$) compounds can be synthesized~\cite{Kawano2003}, 
the latter adopting the same Th$_7$Fe$_3$-type structure as La$_7T_3$ 
superconductors~\cite{Barker2015,Singh2018La7Rh3,Mayoh2021}, which 
frequently exhibit broken TRS in the superconducting state. 
Although selected properties of Re$_3$B and Re$_7$B$_3$ have been 
investigated by different techniques~\cite{Takagiwa2003,Lue2008,Matano2013}, 
their superconducting properties at a microscopic level, in particular, 
the superconducting order parameter, require further investigations. 

In this paper, we report on a comprehensive 
study of the superconducting properties of Re$_3$B and Re$_7$B$_3$, 
carried out via electrical-resistivity, magnetization, heat-capacity, 
and muon-spin relaxation/rotation ($\mu$SR) measurements, as well as by electronic band-structure calculations. 
Our endeavors served a dual purpose. 
Firstly, since La$_7T_3$ shows evidence of TRS breaking below $T_c$, 
it is of interest to establish 
if also the isostructural Re$_7$B$_3$ compound shows similar features. 
Secondly, by systematically investigating the C-Re$_3$B and NC-Re$_7$B$_3$ 
superconductors, the previous findings regarding the Re$T$ family can 
be extended also to other NCSC families, thus providing further insight 
into the open question of TRS breaking in NCSCs.

\begin{figure}[!thp]
	\centering
	\includegraphics[width=0.49\textwidth,angle=0]{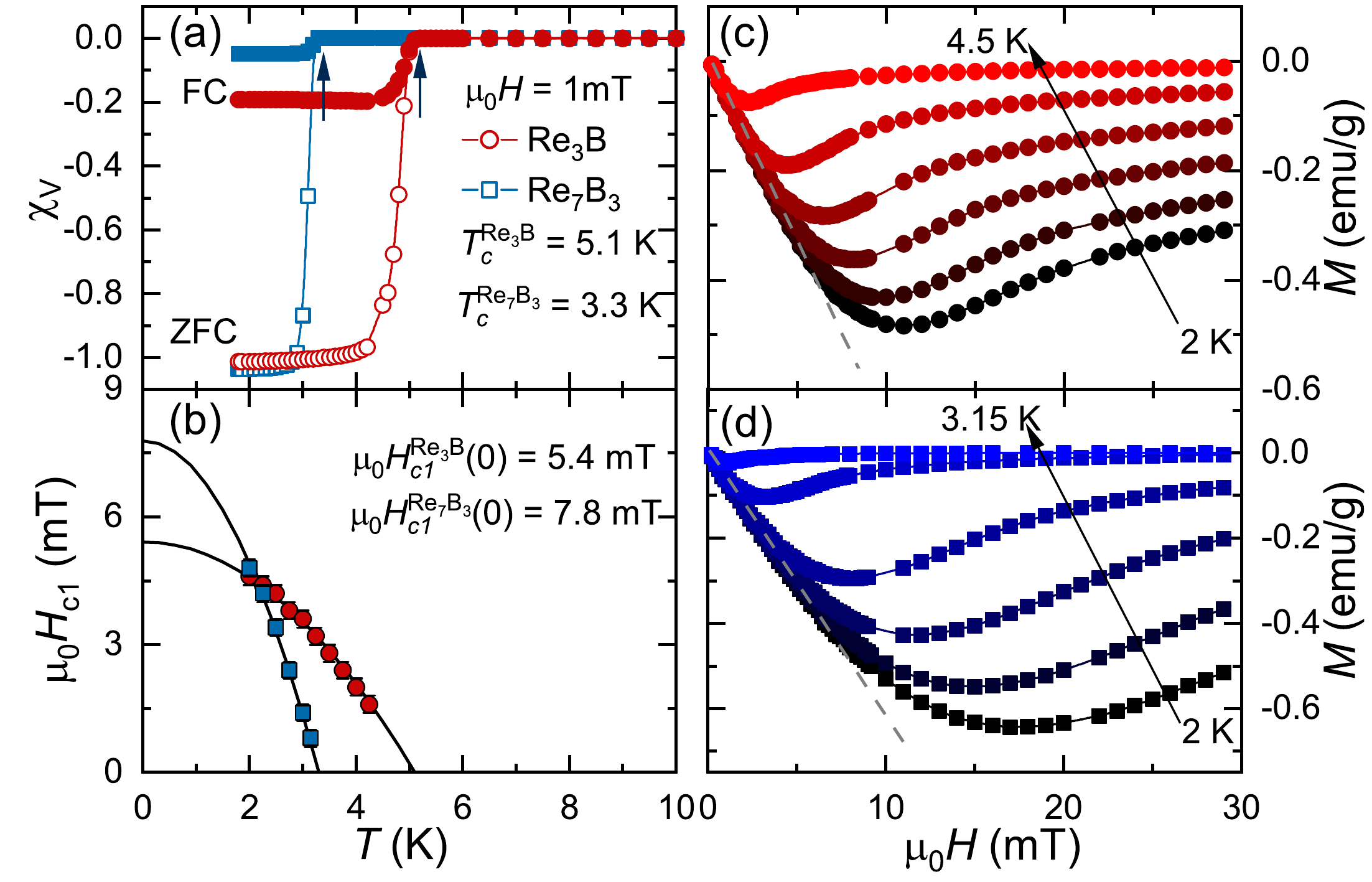}
	\caption{\label{fig:Chi}(a) Temperature-dependent magnetic susceptibility 
		of Re$_3$B and Re$_7$B$_3$, measured in an applied field of 1\,mT 
		using the ZFC and FC protocols.
		(b) Lower critical fields $H_{c1}$ vs.\ temperature. 
		Solid lines are fits to $\mu_{0}H_{c1}(T) =\mu_{0}H_{c1}(0)[1-(T/T_{c})^2]$. 
		Field-dependent magnetization recorded at various temperatures for 
		Re$_3$B (c) and Re$_7$B$_3$ (d). For each temperature, $H_{c1}$ 
		was determined as the value where $M(H)$ starts deviating from 
		linearity (see dashed lines).}
\end{figure}
%

\section{Experimental and numerical methods\label{sec:details}}\enlargethispage{8pt}

Polycrystalline rhenium-boron compounds 
were prepared by arc melting Re (99.99\%, ChemPUR) and B (99.995\%, 
ChemPUR) powders with different stoichiometric ratios in a high-purity 
argon atmosphere. To improve the homogeneity, samples were flipped and remelted several 
times and, finally, annealed at 800$^\circ$C for two weeks. The x-ray 
powder diffraction, measured using a Bruker D8 diffractometer with Cu 
K$\alpha$ radiation, confirmed the orthorhombic centrosymmetric structure 
of Re$_3$B ($Cmcm$, No.\,63), and the hexagonal noncentrosymmetric 
structure of Re$_7$B$_3$ ($P6_3mc$, No.\,186) 
(see details in Fig.~\,S1)~\cite{Supple}. 
The magnetization, electrical-resistivity, and heat-capacity measurements 
were performed on Quantum Design MPMS and PPMS instruments, respectively. 
The bulk $\mu$SR measurements were carried out at the multipurpose 
surface-muon spectrometer (Dolly) of the Swiss muon source at Paul 
Scherrer Institut, Villigen, Switzerland. The $\mu$SR data were analyzed 
by means of the \texttt{musrfit} software package~\cite{Suter2012}.

The electronic band structures of Re$_3$B and Re$_7$B$_3$ were
calculated via the density functional theory (DFT), within the generalized gradient approximation 
(GGA) of Perdew-Burke-Ernzerhof (PBE) realization~\cite{Perdew1996iq}, 
as implemented in the {\sc Quantum ESPRESSO}~\cite{Giannozzi2009,Giannozzi2017}. The projector augmented wave (PAW) pseudopotentials were adopted for the calculation~\cite{Blochl1994zz,Corso2014}. Electrons belonging to the outer atomic configuration 
were treated as valence electrons, here corresponding to 15 electrons in Re ($5s^25p^65d^56s^2$), and 3 electrons in B ($2s^22p^1$). The kinetic energy cutoff was fixed to 55\,Ry. 
For the self-consistent calculation, the Brillouin 
zone integration was performed on a $\Gamma$-centered mesh of 
$15 \times 15 \times 10$ $k$-points for Re$_3$B and $12 \times 12 \times 18$ $k$-points for Re$_7$B$_3$.
Experimentally determined lattice constants and atom positions were used in this calculation.

\section{Results and discussion\label{sec:results}}\enlargethispage{8pt}
The bulk superconductivity of C-Re$_3$B and NC-Re$_7$B$_3$ was first 
characterized by magnetic susceptibility measurements,
using both field-cooled (FC) and zero-field-cooled (ZFC) protocols
in an applied field of 1\,mT. 
As indicated by the arrows in Fig.~\ref{fig:Chi}(a), a clear 
diamagnetic signal appears below the superconducting 
transition at $T_c$ = 5.1\,K and 3.3\,K for Re$_3$B and Re$_7$B$_3$, respectively. 
After accounting for the demagnetization factor, the superconducting shielding fraction of both samples is close to 100\%, 
indicative of bulk SC, which was further confirmed by heat-capacity measurements~\cite{Supple}. 
To determine the lower critical field $H_{c1}$, essential for performing $\mu$SR
measurements on type-II superconductors, the field-dependent magnetization $M(H)$ was 
collected at
various temperatures. Some representative $M(H)$ curves 
are shown in Figs.~\ref{fig:Chi}(c) and (d) for Re$_3$B and  Re$_7$B$_3$, respectively. 
The estimated $H_{c1}$ values as a function of temperature are 
summarized in Fig.~\ref{fig:Chi}(b), resulting in $\mu_{0}H_{c1}(0)$ = 5.4(1)\,mT 
and 7.8(1)\,mT for Re$_3$B and Re$_7$B$_3$, respectively. 

\begin{figure}[htp]
	\centering
	\includegraphics[width=0.48\textwidth,angle= 0]{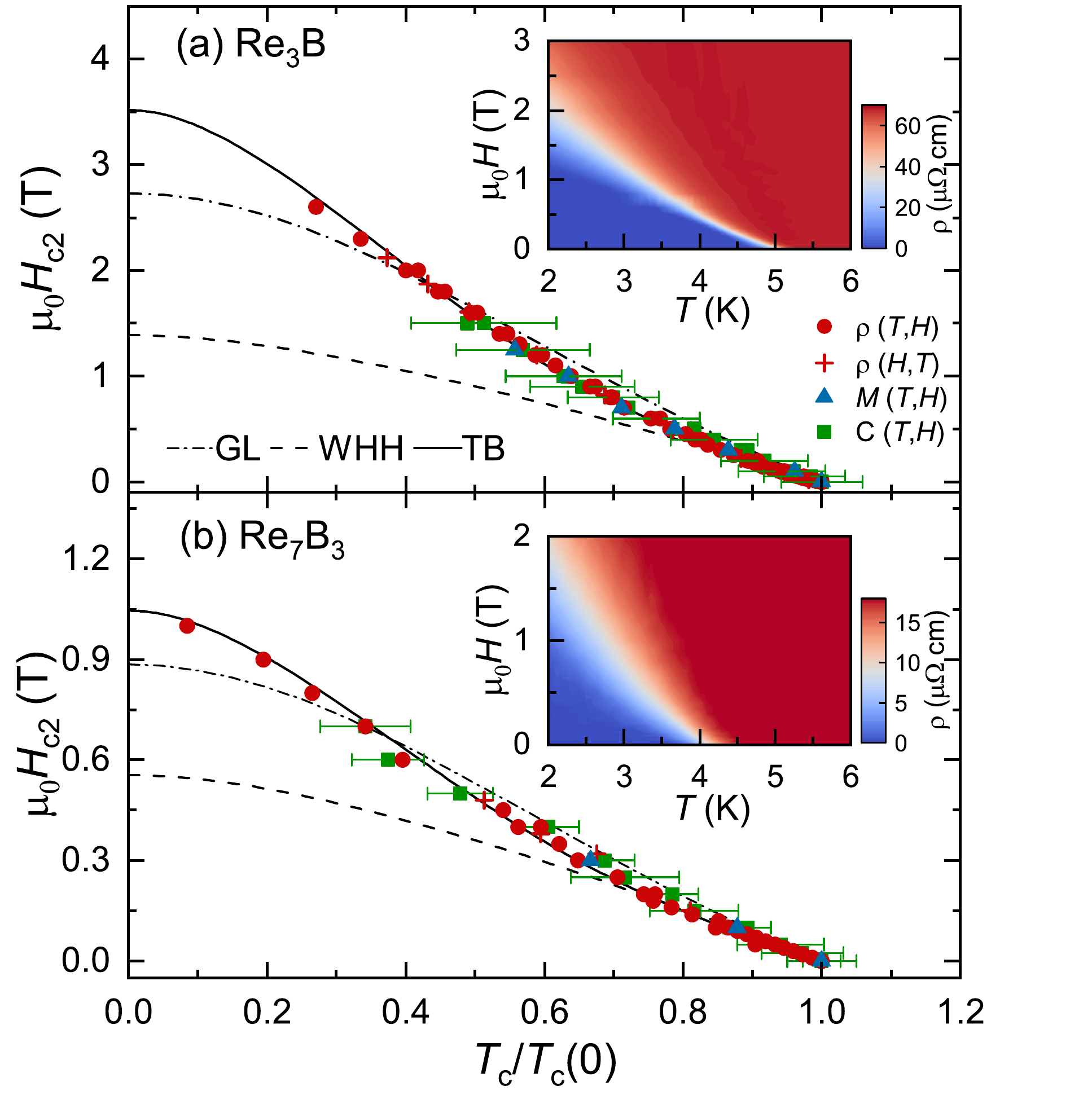}
	\caption{\label{fig:Hc2}Upper critical fields $H_{c2}$ vs.\ reduced 
	temperature $T_c/T_c(0)$ for Re$_3$B (a) and Re$_7$B$_3$ (b), as 
	determined from temperature-dependent electrical resistivity $\rho(T,H)$, 
	magnetization $M(T,H)$, and heat capacity $C(T,H)$, and from field-dependent 
	electrical resistivity $\rho(H,T)$. The contour plots of $\rho(T,H)$ 
	in the insets indicate a clear positive curvature close to $T_c$. Three different 
	models, including the GL- (dash-dotted line), WHH- (dashed line), and TB models (solid line), were used to analyze the $H_{c2}(T)$ data. 
\tcr{Note the positive curvature visible near $\mu_{0}H$ $\sim$ 1\,T and 0.2\,T for Re$_3$B and Re$_7$B$_3$, respectively.} 
	The error bars refer to the superconducting transition widths $\Delta$$T_c$ in the specific-heat data.}
\end{figure}

The upper critical field $H_\mathrm{c2}$ of Re$_3$B and Re$_7$B$_3$ was determined from measurements of the
electrical resistivity, magnetization, and heat capacity under various
magnetic fields up to 3\,T (see Fig.~S2 for details)~\cite{Supple}. 
In zero magnetic field, the $T_c$ values determined from different methods are highly consistent. 
The upper critical fields are summarized in Figs.~\ref{fig:Hc2}(a) and (b) versus the reduced superconducting transition temperature $T_c$/$T_c$(0) 
for Re$_3$B and Re$_7$B$_3$, respectively. The $H_\mathrm{c2}(T)$ was analyzed by
means of Ginzburg–Landau (GL)~\cite{Zhu2008}, Werthamer–Helfand–Hohenberg (WHH)~\cite{Werthamer1966}, and two-band (TB) models~\cite{Gurevich2011}.  
As shown in Fig.~\ref{fig:Hc2}, both GL- and WHH models can reasonably describe the $H_\mathrm{c2}(T)$ at low fields, i.e., $\mu_0H_\mathrm{c2} <$ 0.5\,T (0.2\,T) for Re$_3$B (Re$_7$B$_3$). While at higher magnetic fields, both models deviate significantly from the experimental data and provide underestimated $H_\mathrm{c2}$ values. 
Such discrepancy most likely hints at multiple superconducting gaps in Re$_3$B and Re$_7$B$_3$, as evidenced also by the
positive curvature of $H_\mathrm{c2}(T)$, a typical feature of multigap superconductors, as e.g., Lu$_2$Fe$_3$Si$_5$~\cite{Nakajima2012}, MgB$_2$~\cite{Muller2001,Gurevich2004}, or the recently reported Mo$_5$PB$_2$~\cite{Shang2020MoPB}. Physically, the positive curvature reflects the gradual suppression of the small superconducting gap upon increasing the magnetic field. As clearly demonstrated in the insets of Fig.~\ref{fig:Hc2}, the $H_\mathrm{c2}(T)$ of Re$_3$B and Re$_7$B$_3$ exhibit clear kinks close to 0.5 and 0.2\,T, respectively, most likely coinciding with the field values able to suppress the smaller gap. As shown by the solid lines in Fig.~\ref{fig:Hc2}, the TB model shows a remarkable agreement with the experimental data  and provides 
$\mu_0 H_{c2}(0)$ = 3.5(1)\,T and 1.05(5)\,T for Re$_3$B and Re$_7$B$_3$, respectively.

%
\begin{figure}[!thp]
	\centering
	\includegraphics[width=0.49\textwidth,angle= 0]{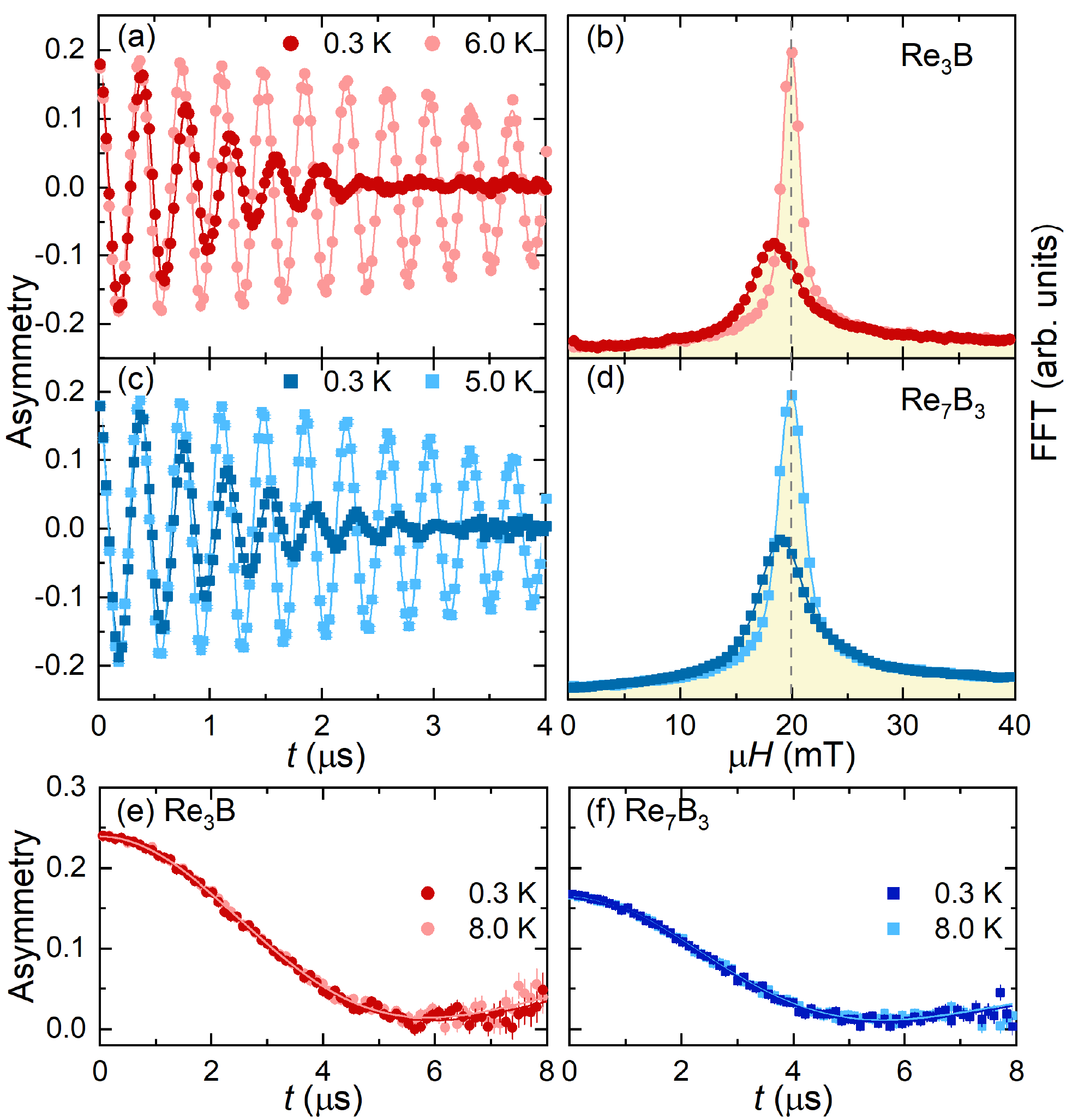}
	\caption{\label{fig:TF-muSR}(a) TF-$\mu$SR spectra of Re$_3$B collected 
	in the superconducting- (0.3\,K) and the normal state (6\,K) in an 
	applied magnetic field of 20\,mT. (b) Fast Fourier transforms of 
	the TF-$\mu$SR data shown in (a). 
	Panels (c) and (d): the analogous results for the Re$_7$B$_3$ case. 
		The solid lines through the data are fits to 
		Eq.~\eqref{eq:TF_muSR}, while the vertical dashed line marks 
		the applied magnetic field. Note the clear diamagnetic shift 
		and the field broadening at 0.3\,K, as shown in (b) and (d).
		ZF-$\mu$SR spectra of Re$_3$B (e) and Re$_7$B$_3$ (f), 
		collected in the superconducting and the normal states. Solid 
		lines are fits using the equation described in the text. The 
		overlapping datasets indicate no evident changes with temperature.}
\end{figure}
\begin{figure}[!thp]
	\centering
	\includegraphics[width=0.49\textwidth,angle= 0]{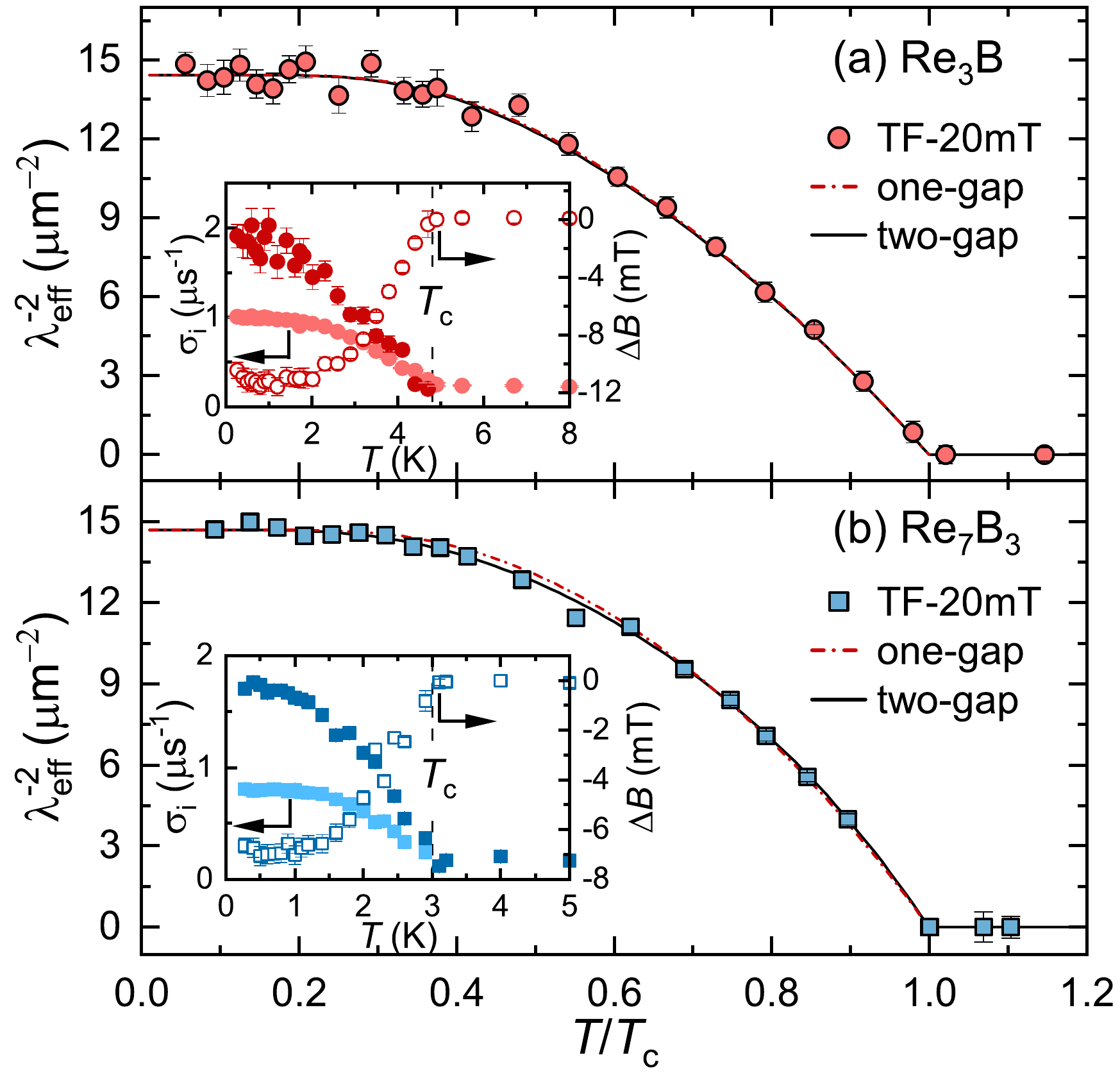}
	\caption{\label{fig:lambda}Superfluid density vs.\ reduced temperature 
	$T/T_c$	for Re$_3$B (a) and Re$_7$B$_3$ (b). The insets show the 
		temperature-dependent muon-spin relaxation rate $\sigma_\mathrm{i}(T)$ (left-axis) 
		and the diamagnetic shift $\Delta B(T)$ (right-axis). The dashed lines in the insets indicate the $T_c$ values, while the dash-dotted- and solid lines in the main panels represent fits to a fully-gapped $s$-wave model with one- and two gaps, respectively.}
\end{figure}

%
%
%
To investigate at a microscopic level the SC of 
Re$_3$B and Re$_7$B$_3$, we carried out systematic transverse-field (TF) 
$\mu$SR measurements in an applied field of 20\,mT, i.e., more than twice 
their $\mu_0H_\mathrm{c1}$(0) values [see Fig.~\ref{fig:Chi}(b)].
Representative TF-$\mu$SR spectra collected in the superconducting- and 
normal states of Re$_3$B and Re$_7$B$_3$ are shown in Figs.~\ref{fig:TF-muSR}(a) and (c), respectively.    
The additional field distribution broadening due to flux-line lattice (FLL) in the mixed state is clearly visible in Figs.~\ref{fig:TF-muSR}(b) and (d), where the fast-Fourier-transform (FFT) spectra of the corresponding TF-$\mu$SR data are presented.
To describe the asymmetric field distribution (e.g., see FFT at 0.3\,K), the TF-$\mu$SR spectra were modeled using:  
\begin{equation}
	\label{eq:TF_muSR}
	A(t) = \sum\limits_{i=1}^n A_i \cos(\gamma_{\mu} B_i t + \phi) e^{- \sigma_i^2 t^2/2} +
	A_\mathrm{bg} \cos(\gamma_{\mu} B_\mathrm{bg} t + \phi).
\end{equation}
Here $A_{i}$, $A_\mathrm{bg}$ and $B_{i}$, $B_\mathrm{bg}$ 
are the asymmetries and local fields sensed by implanted muons in the 
sample and sample holder (copper, which normally shows zero muon-spin depolarization), $\gamma_{\mu}$/2$\pi$ = 135.53\,MHz/T 
is the muon gyromagnetic ratio, $\phi$ is a shared initial phase, and $\sigma_{i}$ 
is the Gaussian relaxation rate of the $i$th component. 
As shown by solid lines in Fig.~\ref{fig:TF-muSR}(a) to (d), two oscillations (i.e., $n = 2$) 
are required to properly describe the TF-$\mu$SR spectra for both Re$_3$B and Re$_7$B$_3$. 
The derived $\sigma_{i}(T)$ as a function of temperature are summarized in the insets of Fig.~\ref{fig:lambda}.
Above $T_c$, $\sigma_{i}(T)$ values are small and temperature-independent, but below $T_c$ they start to increase due to the onset of FLL and the increased
superfluid density. Simultaneously, a diamagnetic field shift appears below $T_c$, given by $\Delta B (T) = \langle B \rangle - B_\mathrm{appl.}$, 
with $\langle B \rangle = (A_1\,B_1 + A_2\,B_2)/A_\mathrm{tot}$, 
$A_\mathrm{tot} = A_1 + A_2$, 
and $B_\mathrm{appl.}$ the applied field.  
The effective Gaussian relaxation rate can be estimated from 
$\sigma_\mathrm{eff}^2/\gamma_\mu^2 = \sum_{i=1}^2 A_i [\sigma_i^2/\gamma_{\mu}^2 + \left(B_i - \langle B \rangle\right)^2]/A_\mathrm{tot}$~\cite{Maisuradze2009}.
Considering the constant nuclear relaxation rate $\sigma_\mathrm{n}$ in 
the narrow temperature range investigated here, confirmed also by zero-field 
(ZF) $\mu$SR measurements [see details in Figs.~\ref{fig:TF-muSR}(e) 
and (f) and Table~S1]~\cite{Supple}, 
the superconducting contribution can be extracted using $\sigma_\mathrm{sc} = \sqrt{\sigma_\mathrm{eff}^{2} - \sigma^{2}_\mathrm{n}}$. Then, the effective magnetic penetrate depth  $\lambda_\mathrm{eff}$ and thus, the superfluid density $\rho_\mathrm{sc}$ ($\propto$ $\lambda_\mathrm{eff}^{-2}$) can be calculated
following $\sigma_\mathrm{sc} = 0.172 \frac{\gamma_{\mu} \Phi_0}{2\pi}(1-h)[1+1.21(1-\sqrt{h})^3]\lambda_\mathrm{eff}^{-2}$~\cite{Brandt2003,Barford1988}, where $h = H_\mathrm{appl}/H_\mathrm{c2}$, with $\mu_0H_\mathrm{appl}$ = 20\,mT the 
applied magnetic field.

We also performed ZF-$\mu$SR measurements in both the normal- and 
the superconducting states of Re$_3$B and Re$_7$B$_3$. As shown in Figs.~\ref{fig:TF-muSR}(e) 
and (f), neither coherent oscillations nor fast decays could be identified in 
the spectra collected above (8\,K) and below $T_c$ (0.3\,K), hence 
implying the lack of any magnetic order or fluctuations. 
The weak muon-spin relaxation in absence of an external magnetic 
field is mainly due to the randomly oriented nuclear moments, which can 
be modeled by means of a phenomenological relaxation function, 
consisting of a combination of Gaussian- and Lorentzian Kubo-Toyabe relaxations~\cite{Kubo1967,Yaouanc2011}, $A(t) = A_\mathrm{s}[\frac{1}{3} + \frac{2}{3}(1 - \sigma_\mathrm{ZF}^{2}t^{2} - \Lambda_\mathrm{ZF} t) \mathrm{e}^{(-\frac{\sigma_\mathrm{ZF}^{2}t^{2}}{2} - \Lambda_\mathrm{ZF} t)}] + A_\mathrm{bg}$.
Here $A_\mathrm{s}$ ($\equiv A_\mathrm{tot}$) and $A_\mathrm{bg}$ are the same as in the TF-$\mu$SR case [see Eq.~\ref{eq:TF_muSR}].
The $\sigma_\mathrm{ZF}$ and $\Lambda_\mathrm{ZF}$ represent the zero-field Gaussian and Lorentzian relaxation rates, respectively.  
As shown by the solid lines in Figs.~\ref{fig:TF-muSR}(e) and (f), the derived relaxations in the normal- and the superconducting states are almost identical (see Table~S1)~\cite{Supple}, as confirmed also by the 
practically overlapping ZF-$\mu$SR 
spectra above and below $T_c$. This lack of evidence for an additional 
$\mu$SR relaxation below $T_c$ excludes a possible TRS breaking in the superconducting state of both C-Re$_3$B and NC-Re$_7$B$_3$.
%

%
%
%
\begin{figure}[th]
	\centering
	\includegraphics[width=0.48\textwidth,angle=0]{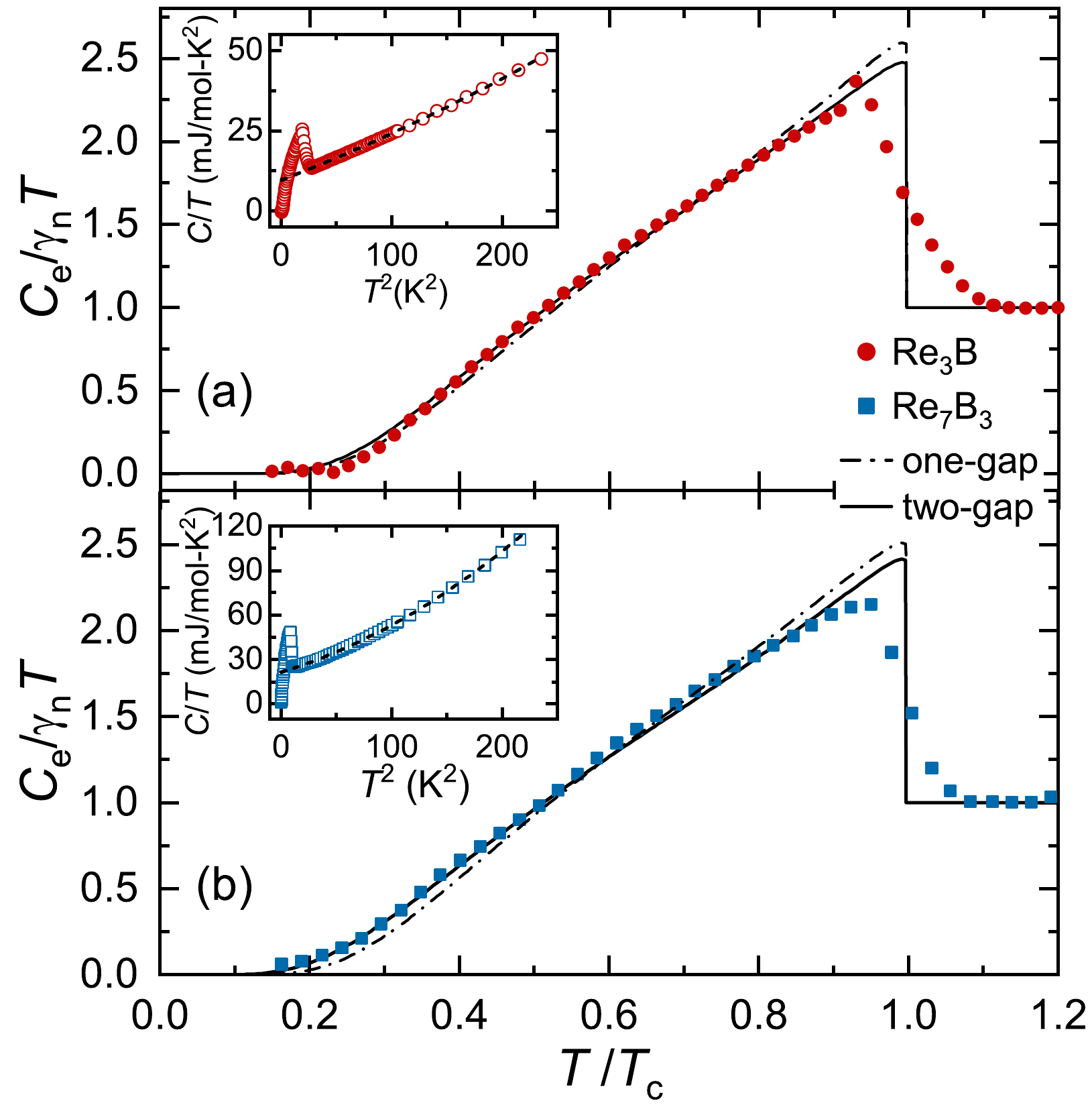}
	\caption{\label{fig:Cp}%
		Normalized electronic specific heat $C_\mathrm{e}/\gamma_n T$ versus reduced temperature $T/T_c$ for Re$_3$B (a) and Re$_7$B$_3$ (b). 
		$\gamma_n$ is the normal-state electronic specific-heat coefficient. The insets show the measured specific heat $C/T$ versus $T^2$. The 
		dashed-lines in the insets are fits to $C/T = \gamma_n + \beta T^2 + \delta T^4$ for $T > T_c$, while the dash-dotted- and solid lines in the main panel represent the  
		electronic specific heat calculated by considering a fully-gapped $s$-wave model with one- and two gaps, respectively.} 
\end{figure}
%
%
The superfluid density $\rho_\mathrm{sc}$ vs.\ the reduced $T/T_c$ are shown 
in Figs.~\ref{fig:lambda}(a) and (b) for Re$_3$B and Re$_7$B$_3$, respectively. 
The temperature-independent superfluid density below $T_c/3$ indicates a 
fully-gapped SC in both cases. Therefore, we analyzed the $\rho_\mathrm{sc}(T)$ 
by means of a fully-gapped $s$-wave model: 
\begin{equation}
	\label{eq:rhos}
	\rho_\mathrm{sc}(T) = \frac{\lambda_\mathrm{eff}^{-2}(T)}{\lambda_0^{-2}} = 1 + 2\int^{\infty}_{\Delta(T)} \frac{\partial f}{\partial E} \frac{EdE}{\sqrt{E^2-\Delta^2(T)}}.
\end{equation}
Here $f = (1+e^{E/k_\mathrm{B}T})^{-1}$ and $\Delta(T)$ are the Fermi- and the su\-per\-con\-duc\-ting\--gap functions. 
The $\Delta(T)$ is assumed to follow  $\Delta(T) = \Delta_0 \mathrm{tanh} \{ 1.82[1.018(T_\mathrm{c}/T-1)]^{0.51} \}$ 
\cite{Carrington2003}, where $\Delta_0$ is the superconducting gap at 0\,K.
Since the upper critical field $H_\mathrm{c2}$(T) exhibits typical features of multigap SC (see Fig.~\ref{fig:Hc2}), the superfluid density was fitted using Eq.~\eqref{eq:rhos} with one- and two gaps. In the two-gap case, $\rho_\mathrm{sc}(T) = w \rho_\mathrm{sc}^{\Delta^\mathrm{f}}(T) + (1-w) \rho_\mathrm{sc}^{\Delta^\mathrm{s}}(T)$, with $\rho_\mathrm{sc}^{\Delta^\mathrm{f}}$ and 
$\rho_\mathrm{sc}^{\Delta^\mathrm{s}}$ being the superfluid densities related to 
the first- ($\Delta^\mathrm{f}$) and second ($\Delta^\mathrm{s}$) gap, and $w$ a relative weight. 
For Re$_3$B, both the one- and two-gap models show an almost identical goodness-of-fit parameter ($\chi^2_{r} \sim 1.2$), reflected in two practically overlapping fitting curves 
in Fig.~\ref{fig:lambda}(a). 
For Re$_7$B$_3$, instead, the two-gap model ($\chi^2_{r} \sim 1.1$) is slightly 
superior to the one-gap model ($\chi^2_{r} \sim 2.2$) [see Fig.~\ref{fig:lambda}(b)]. 
For the two-gap model, in the Re$_3$B case, the derived zero-temperature magnetic penetration depth is $\lambda_\mathrm{0} = 263(2)$\,nm, the gap values are $\Delta_0^\mathrm{f}$ = 0.72(1)\,meV and  $\Delta_0^\mathrm{s}$ = 0.87(2)\,meV, with a weight $w$ = 0.7. In the Re$_7$B$_3$ case, the corresponding 
values are $\lambda_\mathrm{0}$ =261(2)\,nm, $\Delta_0^\mathrm{f}$ = 0.35(1)\,meV and  $\Delta_0^\mathrm{s}$ = 0.57(2)\,meV, with $w$ = 0.27. 
For the one-gap model, the gap values are $\Delta_0 = 0.77(2)$ and 0.50(2)\,meV for Re$_3$B and Re$_7$B$_3$, with the same $\lambda_\mathrm{0}$ values as in the two-gap case.

Unlike in the clean-limit case ($\xi_0$ $\ll$ $l_\mathrm{e}$) [see Eq.~\eqref{eq:rhos}], 
in the dirty limit, the BCS coherence length $\xi_0$ is much larger than 
the electronic mean-free path $l_\mathrm{e}$. In the BCS approximation, 
the temperature dependent superfluid density in the dirty limit is given by~\cite{Tinkham1996}:
\begin{equation}
		\label{eq:dirty}
		\rho_\mathrm{sc}(T) = \frac{\Delta(T)}{\Delta_0} \mathrm{tanh} \left[\frac{\Delta(T)}{2k_\mathrm{B}T}\right],
\end{equation}
where $\Delta(T)$ is the same as in Eq.~\eqref{eq:rhos}.
For Re$_3$B, $\xi_0$ is larger than $l_\mathrm{e}$ ($\xi_0$/$l_\mathrm{e}$ $\sim$ 7.7), 
therefore, Re$_3$B is close to the dirty limit; while for Re$_7$B$_3$, $\xi_0$ 
is smaller than $l_\mathrm{e}$ ($\xi_0$/$l_\mathrm{e}$ $\sim$ 0.3), 
i.e., it is close to the clean limit.  
For both compounds, the $\xi_0$ and $l_\mathrm{e}$ are not significantly  
different and exhibit similar magnitudes. Hence, both Eq.~\eqref{eq:rhos} 
and Eq.~\eqref{eq:dirty} describe quite well the low-$T$ superfluid density, and yield similar superconducting gaps (see Table~\ref{tab:parameter}).	

To further support the indications of a multigap SC obtained from $H_{c2}$,  
we measured also the zero-field specific heat down to 1/3$T_c$. 
After subtracting the phonon contribution ($\beta$$T^2$ + $\delta$$T^4$) from the measured data, the obtained electronic specific
heat divided by the $\gamma_\mathrm{n}$, i.e., $C_\mathrm{e} / \gamma_\mathrm{n} T$, is shown in Fig.~\ref{fig:Cp}(a) and (b) vs.\ the reduced temperature $T/T_c$ for Re$_3$B and Re$_7$B$_3$, respectively. 
The superconducting-phase contribution to the entropy can be calculated following the BCS expression~\cite{Tinkham1996}:
\begin{equation}
	\label{eq:entropy}
	S(T) = -\frac{6\gamma_\mathrm{n}}{\pi^2 k_\mathrm{B}} \int^{\infty}_0 [f\mathrm{ln}f+(1-f)\mathrm{ln}(1-f)]\,\mathrm{d}\epsilon,
\end{equation}
where $f$ is the same as in Eq.~\eqref{eq:rhos}. Then, the temperature-dependent electronic specific heat below $T_c$ can be obtained from $C_\mathrm{e}(T) =T \frac{dS(T)}{dT}$.
The dash-dotted lines in Fig.~\ref{fig:Cp} represent fits of an $s$-wave model with $\gamma_\mathrm{n}$ = 9.6(1) and 21.5(2) mJ/mol-K$^2$ and a single gap $\Delta_0$ = 0.75(2) and 0.47(1)\,meV for Re$_3$B and Re$_7$B$_3$, respectively.
\tcr{For Re$_7$B$_3$, while the one-gap model reproduces the data for $T/T_c \gtrsim 0.5$, it deviates from them at lower temperatures, hence yielding a slightly larger $\chi^2_{r} \sim$ 7.8 than the two-gap model (see below).}   
On the contrary, the two-gap model exhibits a better agreement with the experimental data. The solid line in Fig.~\ref{fig:Cp}(b) is a fit with two energy gaps, i.e., $C_e(T)/T = wC_e^{\Delta^\mathrm{f}}(T)/T + (1-w)C_e^{\Delta^\mathrm{s}}(T)/T$.  
Here, each term represents a one-gap specific-heat contribution, 
with $w$, $\Delta^\mathrm{f}$, and $\Delta^\mathrm{s}$ being the same 
parameters as in case of superfluid density fits.   
\tcr{
In Re$_7$B$_3$, by sharing the $w$ values, the two-gap model gives 
$\Delta_0^\mathrm{f} = 0.32(1)$\,meV and $\Delta_0^\mathrm{s} = 0.50(1)$\,meV, 
with $\chi^2_{r} \sim \tcr{1.7}$.} 
\tcr{Similarly, in Re$_3$B, the solid-line in Fig.~\ref{fig:Cp}(a) is a fit with 
$\Delta_0^\mathrm{f} = 0.69(2)$\,meV and $\Delta_0^\mathrm{s} = 0.79(2)$\,meV. 
For Re$_3$B, although the two-gap model agrees slightly better with 
the experimental data for $T/T_c > 0.4$, below it, both one-gap and two-gap models 
deviate slightly from the measured data, probably reflecting an improper 
subtraction of the nuclear Schottky contribution (due to the limited lowest 
temperature that could be reached in this study --- see details in 
Fig.~S3)~\cite{Supple}. Measurements of zero-field specific heat down to 
the mK range are highly demanded to confirm the multigap nature of 
Re$_3$B and Re$_7$B$_3$.} 
\tcr{Note that, for both compounds,} the superconducting parameters 
determined from the specific heat and the TF-$\mu$SR are highly consistent (see Table~\ref{tab:parameter}). 


\begin{figure}[th]
	\centering
	\includegraphics[width=0.48\textwidth,angle=0]{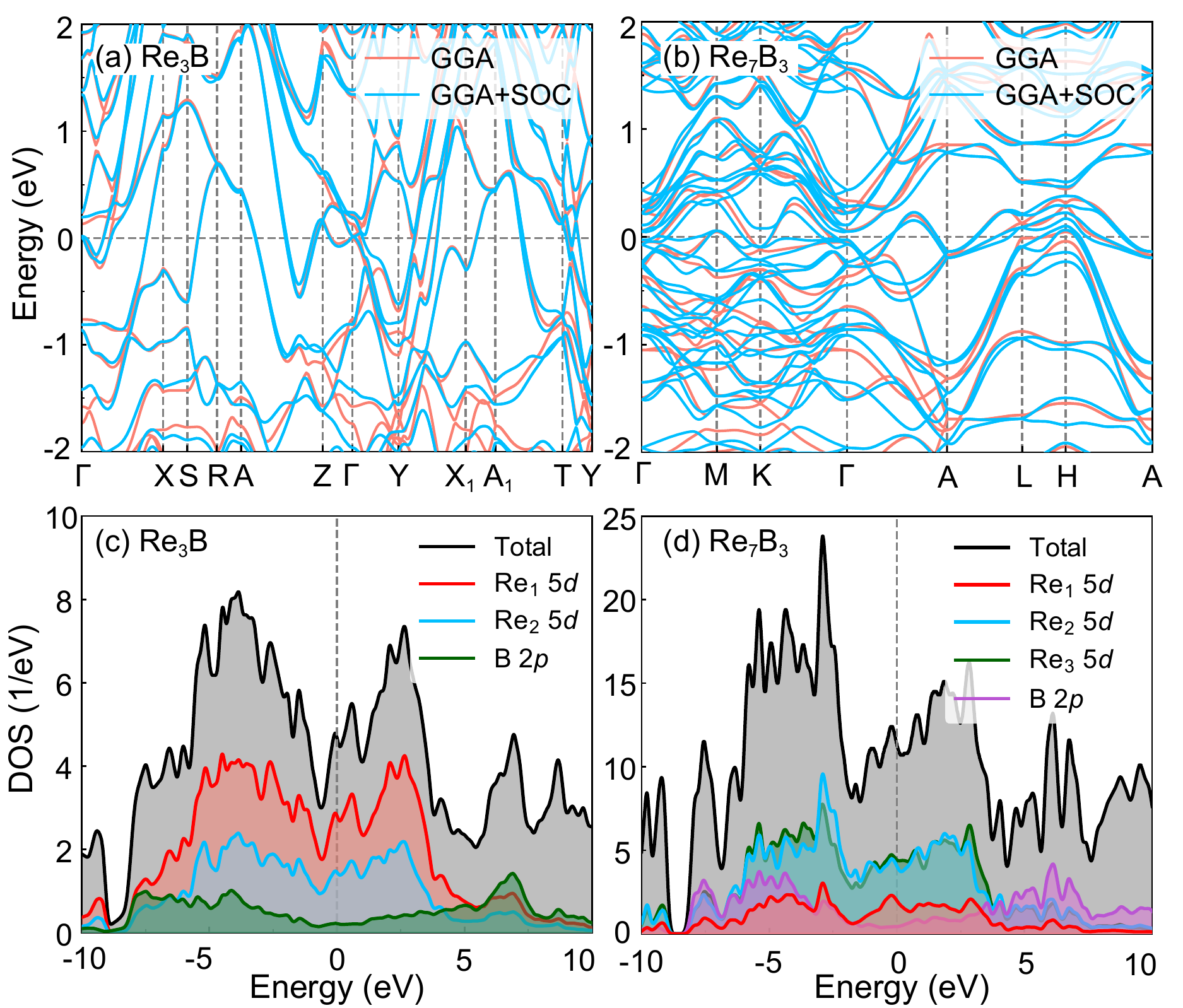}
	\caption{\label{fig:dft} Electronic band structure of C-Re$_3$B (a) 
	and NC-Re$_7$B$_3$ (b), calculated by ignoring (red) and by 
	considering (blue) the spin-orbit coupling. For both compounds, 
	various bands which cross the Fermi level can be identified. 
	The total and partial (Re $5d$ and B $2p$ orbitals) density of states are shown in panels (c) and (d) for Re$_3$B and Re$_7$B$_3$, respectively.}
\end{figure}
%

Further evidence on the multigap SC and insight into the electronic 
properties of Re$_3$B and Re$_7$B$_3$ comes from
band-structure calculations.
The electronic band structures, as well as the density of states (DOS) 
are shown in Fig.~\ref{fig:dft}.
As can be seen from Figs.~\ref{fig:dft}(a) and (b), 4 and 12 different 
bands cross
the Fermi level in Re$_3$B and Re$_7$B$_3$, respectively. 
Close to $E_\mathrm{F}$, the DOS of both compounds is dominated by 
the Re 5$d$ orbitals, while the contribution from the B $2p$ orbitals is 
negligible [see Figs.~\ref{fig:dft}(a) and (b) and Fig.~S4~\cite{Supple}]. 
Away from the Fermi level, the Re $5d$ and B $2p$ orbitals are highly 
hybridized. The estimated DOS at $E_\mathrm{F}$
are $\sim 4.7$ and $\sim 11.4$\,states/(eV$\cdot$u.c.) for Re$_3$B and 
Re$_7$B$_3$, both comparable to the experimental values determined from electronic specific heat coefficient (see Table~\ref{tab:parameter}). 
We expect the multigap features of
Re$_3$B and Re$_7$Be$_3$, to be closely related to the different site symmetries of Re atoms in the unit cell. For Re$_3$B, according to band-structure calculations, the contribution of Re1 (8$f$) atoms to the DOS is comparable to that of Re2 (4$c$) atoms [see Figs.~\ref{fig:dft}(c)]. While for Re$_7$B$_3$, the contribution of Re2 (6$c$) and Re3 (6$c$) atoms is preponderant compared to that of Re1 (2$b$) atoms.

Now, let us compare the superconducting parameters of 
Re$_3$B and Re$_7$B$_3$ with those of other superconductors. 
First, by using the SC parameters obtained from the measurements presented here, 
we calculated an effective Fermi temperature $T_\mathrm{F}$ = 1.0(1)$\times$10$^4$ 
and 0.25(1)$\times$10$^4$\,K for Re$_3$B and Re$_7$B$_3$ (see other parameters in Table~\ref{tab:parameter}). $T_\mathrm{F}$ is proportional to $n_\mathrm{s}^{2/3}$/$m^\star$, where $n_\mathrm{s}$ and $m^\star$ are the carrier density and the effective mass. Consequently, 
the different families of superconductors can be classified according to 
their $T_c$/$T_\mathrm{F}$ ratios into a so-called Uemura plot~\cite{Uemura1991}. Several types of unconventional superconductors, including heavy fermions, organics, high-$T_c$ iron pnictides, and cuprates, all lie in a 10$^{-2}$ $< T_c/T_\mathrm{F} <$ 10$^{-1}$ band (gray region in Fig.~S5)~\cite{Supple}. Conversely, many conventional superconductors, e.g., Al, Sn, and Zn, are located at $T_c/T_\mathrm{F} < \sim$10$^{-4}$. Between these two categories lie 
several multigap superconductors, e.g., LaNiC$_2$, Nb$_5$Ir$_3$O, ReBe$_{22}$, NbSe$_2$, and MgB$_2$~\cite{Shang2019c,Uemura1991,Uemura2006,Shang2020NbIr,Shang2019b}. 
Although there is no conclusive 
evidence for them to be classified as unconventional superconductors, 
the rhenium-boron superconductors lie clearly far off the conventional band. For Re$_3$B, 
$T_c$/$T_\mathrm{F}$ = 4.8$\times$10$^{-4}$ is almost identical to the 
analogous value for 
multigap ReBe$_{22}$ and LaNiC$_2$, the latter 
representing a typical example of NCSCs. 
While for Re$_7$B$_3$, the $T_c$/$T_\mathrm{F}$ = 1.16$\times$10$^{-3}$ is very close to the multigap Nb$_5$Ir$_3$O and elementary rhenium 
superconductors~\cite{Shang2018b,Shang2020ReMo,Shang2020NbIr}, the latter showing a breaking of TRS in the superconducting state and exhibiting a centrosymmetric crystal structure. In general, most of the weakly-correlated NCSCs, e.g., Re$T$, Mo$_3$Al$_2$C, Li$_2$(Pd, Pt)$_3$B, and LaNiC$_2$, exhibit a $T_c/T_\mathrm{F}$ between the unconventional and conventional bands~\cite{Shang2019b}, which is also the case for Re$_7$B$_3$.

Second, we discuss why the multigap feature is more evident in Re$_7$B$_3$ than in Re$_3$B, both in the temperature-dependent superfluid density and the zero-field electronic specific data. In general, if the weight of  
the second gap is relatively small and the gap sizes are not significantly different, this makes it difficult to discriminate between a single- and a two-gap superconductor based on temperature-dependent superconducting properties. 
For Re$_3$B, the weight of the second gap $w$ = 0.3 is similar to that of Re$_7$B$_3$ ($w$ = 0.27). However, the gap sizes are clearly distinct in Re$_7$B$_3$ ($\Delta^\mathrm{f}/\Delta^\mathrm{s}$ $\sim$ 0.6) compared to Re$_3$B ($\Delta^\mathrm{f}/\Delta^\mathrm{s}$ $\sim$ 0.9).
As a consequence, the multigap feature is more evident in Re$_7$B$_3$. On the other hand, 
from the analysis of $H_{c2}(T)$ using a two-band model, the derived inter-band 
and intra-band couplings are $\lambda_{12}$ = 0.08 and $\lambda_{11} \sim \lambda_{22}$ = 0.4, and $\lambda_{12}$ = 0.01 and $\lambda_{11}$ $\sim$ $\lambda_{22}$= 0.15 for Re$_3$B and Re$_7$B$_3$, respectively. In both cases, the inter-band coupling is much smaller than the intra-band coupling.  
In addition, the inter-band coupling of Re$_3$B (0.08) is larger 
than that of Re$_7$B$_3$ (0.01). This makes the gaps to open at 
different electronic bands, less distinguishable in the former case~\cite{Kogan2009}.
Despite these differences, the underlying multigap SC feature of 
both samples is 
reflected in their upper critical fields $H_\mathrm{c2}(T)$ (see Fig.~\ref{fig:Hc2}). To get further insight into the multigap SC of Re$_3$B and Re$_7$B$_3$, 
the measurement of the \emph{field}-dependent superconducting Gaussian relaxation rate $\sigma_\mathrm{sc}(H)$ and of the electronic specific
heat coefficient $\gamma(H)$ provides a possible alternative, both datasets being expected to show a distinct field response compared to a single-gap superconductor~\cite{Shang2019c,Shang2020MoPB}. For example, $\gamma(H)$ exhibits a clear change in slope when the applied magnetic field suppresses the small gap, a feature recognized as the fingerprint of multigap superconductors.
Conversely, $\gamma(H)$ is mostly linear in the single-gap case.

%
\begin{table}[!th]
	\centering
	\caption{Normal- and superconducting state properties of C-Re$_3$B and NC-Re$_7$B$_3$, as 
			determined from electrical-resistivity, magnetization, specific-heat, and $\mu$SR measurements, 
			as well as electronic band-structure calculations. The London penetration depth $\lambda_\mathrm{L}$, the effective mass $m^{\star}$, carrier density $n_\mathrm{s}$, BCS coherence length $\xi_0$,
			electronic mean-free path $l_\mathrm{e}$, Fermi velocity $v_\mathrm{F}$, and effective Fermi temperature $T_\mathrm{F}$ were estimated following the methods in Ref.~\onlinecite{Shang2020ReMo}.
		    \label{tab:parameter}}
	\begin{ruledtabular}
		\begin{tabular}{lccc}
			Property                                 & Unit               & Re$_3$B       & Re$_7$B$_3$    \\ \hline
			Space group                              & --                  & $Cmcm$       & $P6_{3}mc$     \\
			Inversion center                         & --                  & Yes          & No             \\
			$\rho_0$                                 & $\mu\Omega$ cm     & 68.0          & 18.5           \\
			Residual resistivity ratio                                      & --                  & 1.9          & 5.8            \\
			$T_c^\rho$                               & K                  & 5.2           & 3.5            \\
			$T_c^\chi$                               & K                  & 5.1           & 3.3            \\
			$T_c^C$                                  & K                  & 4.7           & 3.1            \\
			$T_c^{\mu\mathrm{SR}}$                   & K                  & 4.8           & 2.9            \\
			$\mu_0H_{c1}$                            & mT                 & 5.4(1)        & 7.8(1)         \\
			$\mu_0H_{c2}$                            & T                  & 3.5(1)        & 1.05(5)        \\
			$\gamma_n$                               & mJ/mol-K$^2$       & 9.6(1)        & 21.5(2)        \\
			$\Theta_\mathrm{D}$                      & K                  & 390(3)        & 440(5)         \\
			$N(\epsilon_\mathrm{F})^C$               & states/eV-f.u.     & 4.1(1)        & 9.1(1)         \\
			$N(\epsilon_\mathrm{F})^\mathrm{DFT}$    & states/eV-f.u.     & 2.35          & 5.7            \\ %
			$\Delta_0(C)$                            & meV                & 0.75(2)       & 0.47(1)        \\
			$\Delta_0$($\mu\mathrm{SR}$)$^\mathrm{clean}$             & meV               & 0.77(2)        & 0.50(2)           \\  
			$\Delta_0$($\mu\mathrm{SR}$)$^\mathrm{dirty}$             & meV               & 0.66(2)        & 0.44(1)           \\  
			$w$                                      & --                  & 0.7          & 0.27           \\
			$\Delta_0^f(C)$                          & meV                & 0.69(2)       & 0.32(1)           \\  
			$\Delta_0^s(C)$                          & meV                & 0.79(2)       & 0.50(1)           \\  
			$\Delta_0^f(\mu\mathrm{SR})$             & meV                & 0.72(1)       & 0.35(1)           \\  
			$\Delta_0^s(\mu\mathrm{SR})$             & meV                & 0.87(2)       & 0.57(2)           \\  
			$\lambda_0$                              & nm                 & 263(2)        & 261(2)            \\[2mm]
			$\lambda_\mathrm{GL}(0)$                 & nm                 & 353(4)        & 259(2)            \\
			$\xi(0)$                                 & nm                 & 9.7(1)        & 17.7(4)           \\
			$\kappa$                                 & --                  & 36(1)        & 14.6(5)           \\
			$\lambda_\mathrm{L}$                     & nm                 & 90(5)         & 229(2)            \\
			$l_\mathrm{e}$                           & nm                 & 2.2(1)        & 22(1)             \\
			$\xi_0$                                  & nm                 & 17(1)         & 6.4(1)             \\
		   	$\xi_0$/$l_\mathrm{e}$                   & --                 & 7.7           & 0.3                \\
			$m^{\star}$                              & $m_e$              & 7.0(2)        & 10.4(1)            \\
			$n_\mathrm{s}$                           & 10$^{28}$\,m$^{-3}$ & 2.4(3)       & 0.56(1)            \\
			$v_\mathrm{F}$                           & 10$^5$\,ms$^{-1}$   & 1.5(1)       & 0.61(1)            \\
			$T_\mathrm{F}$                           & 10$^4$\,K           & 1.0(1)       & 0.25(1)            \\
		\end{tabular}	
	\end{ruledtabular}
\end{table}
%

Finally, we discuss about the effects of a lack of inversion symmetry in 
Re$_7$B$_3$. In NCSCs, the antisymmetric spin-orbit coupling (ASOC) 
allows for the occurrence of an admixture of singlet and triplet pairings, whose
mixing degree is generally believed to be related to the strength of the ASOC~\cite{Bauer2012} and, thus, to unconventional SC. 
Here by comparing NC-Re$_7$B$_3$ with C-Re$_3$B, we found that a noncentrosymmetric structure and its accompanying ASOC have little effect on 
the superconducting properties of Re$_7$B$_3$. First, the upper critical field of NC-Re$_7$B$_3$ is three times smaller than that of C-Re$_3$B, in both cases $H_\mathrm{c2}$ being well below the Pauli limit. Second, according to the ZF-$\mu$SR data (Fig.~\ref{fig:TF-muSR}), TRS is preserved in the superconducting states of both samples. The new results presented here further support the idea that the rhenium presence and its amount are the two key factors which determine the appearance of TRS breaking in Re-based superconductors, while the noncentrosymmetric structure plays only a marginal role. 
Obviously, the Re-content in both Re$_3$B and Re$_7$B$_3$ might be below the threshold value, e.g., 88\% in Re-Mo alloys~\cite{Shang2020ReMo}.
Therefore, it could be interesting to check, 
upon increasing the Re-content, whether the TRS breaking effect will appear 
also in the rhenium-boron superconductors. Third, both Re$_7$B$_3$ and Re$_3$B exhibit nodeless SC with multiple gaps. In case of Re$_7$B$_3$, whether the multigap feature 
is due to the band splitting caused by the ASOC, or to the multiple bands crossing its Fermi level (the latter, in principle, accounting also for the C-Re$_3$B case), requires further theoretical work. 
Overall, as can be seen from Fig.~\ref{fig:dft}(b), the ASOC and the band splitting is relatively small in Re$_7$B$_3$. 
Hence, we expect the spin-singlet pairing to be dominant in both 
the centrosymmetric and noncentrosymmetric rhenium-boron superconductors.

\section{\label{ssec:Sum}Conclusion}\enlargethispage{8pt}
To summarize, we studied the superconducting properties of the centrosymmetric 
Re$_3$B and the noncentrosymmetric Re$_7$B$_3$ superconductors by means of
electrical resistivity, magnetization, heat capacity, and $\mu$SR techniques, 
as well as via numerical calculations. 
The superconducting state of Re$_3$B and Re$_7$B$_3$ is characterized by $T_c$ = 5.1\,K and 3.3\,K, and upper critical fields $\mu_0H_{c2}(0)$ = 3.5\,T and 1.05\,T, respectively. 
The temperature-dependent zero-field electronic specific heat and superfluid density reveal a \emph{nodeless} superconductivity, well described by an \emph{isotropic $s$-wave} model.
\tcr{Both Re$_3$B and Re$_7$B$_3$ exhibit a positive curvature
in their tem\-per\-a\-ture\--de\-pen\-dent upper critical field $H_{c2}(T)$, 
an established fingerprint 
of multigap superconductivity.} 
By combining our extensive experimental results with numerical band-structure calculations, 
we provide 
evidence regarding multigap superconductivity 
in both centro- and noncentrosymmetric rhenium-boron superconductors. 
Finally, the lack of spontaneous magnetic fields below $T_c$ indicates 
that, unlike in the Re$T$ or elementary rhenium, the time-reversal symmetry 
is \emph{preserved} in the superconducting state of both Re$_3$B and Re$_7$B$_3$.
Our results suggest that the spin-singlet paring channel is dominant 
in the rhenium-boron superconductors. 

\emph{Note}. While preparing the current manuscript, we became 
aware of a related work by S.\ Sharma {et al.}~\cite{Sharma2020}, 
in which similar compounds were studied via the $\mu$SR technique.

\begin{acknowledgments}
This work was supported by start funding from 
East-China Normal University (ECNU), the Swiss SNF Grants (No. 200021-169455 and No. 206021-139082) and the Sino-Swiss Science and Technology Cooperation (Grant
No. IZLCZ2-170075).
H.Q.Y.\ acknowledges support from the National Key R\&D Program of China (No.\ 2017YFA0303100 and No.\ 2016YFA0300202), the Key R\&D Program of Zhejiang Province, China (No. 2021C01002), the National Natural Science Foundation of China (No.\ 11974306). We acknowledge the allocation of beam time at the Swiss muon source, and thank the scientists of Dolly $\mu$SR spectrometer for their support.
\end{acknowledgments}

\bibliography{ReB.bib}

\end{document}